\documentclass[prl,twocolumn,superscriptaddress,amsmath,amssymb]{revtex4}

\usepackage{graphicx}
\usepackage{dcolumn}
\usepackage{bm}
\usepackage{epsfig}
\newcommand{\be}{\begin{equation}}
\newcommand{\ee}{\end{equation}}

\begin{document}

\title{Nonequilibrium glass transitions in driven and active matter}

\author{Ludovic Berthier}
\affiliation{Laboratoire Charles Coulomb,
UMR 5221 CNRS and Universit\'e Montpellier 2, Montpellier, France}

\author{Jorge Kurchan}
\affiliation{PMMH, UMR 7636 CNRS and ESPCI, Paris, France}

\date{\today}
\begin{abstract}
The glass transition, extensively studied in
dense fluids, polymers, or colloids,
corresponds to a dramatic evolution of equilibrium transport
coefficients upon a modest change of control 
parameter, like temperature or pressure. A similar phenomenology 
is found in many systems evolving 
far from equilibrium, such as driven granular media, active 
and living matter.   
While many theories compete to describe the glass transition 
at thermal equilibrium, 
very little is understood far from equilibrium.
Here, we solve the dynamics of a specific, yet 
representative, class of glass models 
in the presence of nonthermal driving forces and energy dissipation, 
and show that a dynamic arrest can take place in 
these nonequilibrium conditions.
While the location of the transition 
depends on the specifics of the driving mechanisms, important features 
of the glassy dynamics are insensitive to details, suggesting that 
an `effective' thermal dynamics generically emerges at long time scales
in nonequilibrium systems close to dynamic arrest.
\end{abstract}



\maketitle

Our goal is to study theoretically whether 
a glass transition can occur in a many-body interacting system
driven by fluctuations which do not have a thermal origin. 
We motivate this very general question using 
recent experimental findings, which we divide into two 
broad classes. 
First, in the presence of mechanical 
forcing, granular media display a form of dynamical
slowing down which shares important similarities with the 
dynamics of molecular liquids near the glass transition. 
Forces driving the dynamics of the grains can for instance be a global 
oscillatory shear~\cite{marty}, high-frequency vertical 
vibrations~\cite{reis,tanaka}, or an air flow through the granular 
assembly~\cite{durian}.  The injected mechanical energy 
produces erratic grain motions, while energy is dissipated through 
collisions. Dense granular fluids driven through dynamic arrest 
are also studied numerically~\cite{zippelius}. 
Active and living materials, such as self-propelled colloids
and grains~\cite{lyd,lyd2,deseigne}, cells and 
bacteria~\cite{weitz,petitjean} form a second class of 
nonequilibrium materials where glassy effects 
are reported and a dynamic arrest is observed at large 
density. Numerical studies of simple 
models of active matter in the dense regime have recently 
appeared~\cite{mossa,mossa2,marchetti,loewen},
similarities with the equilibrium 
glass problem were noted~\cite{weitz}.
More generally, the issue of mapping the far-from-equilibrium
dynamics of active and self-propelled particles to an `effective' 
equilibrium problem is a recurring 
theoretical question~\cite{loewen,tailleur,tailleur2,reviewmarchetti}, 
for which we are able to provide an explicit solution in the particular
context of the dynamics of driven glassy materials.

Previous investigations either assume from the start, or 
demonstrate through detailed observations, that the
observed dramatic change in dynamic properties is `similar' 
to the glass transition of simple fluids at thermal 
equilibrium~\cite{rmp}, 
in the
sense that typical signatures of glassy dynamics 
(caging effect, two-step decay of time correlation functions, 
dynamic heterogeneity, etc.) are found.
Theoretical studies exist for particular 
systems~\cite{zippelius,wolynes,mossa}, and this broad set of 
experimental observations raises, we believe,
a more general question that was not adressed in previous work.
Can a glass transition occur far from thermal equilibrium, and how 
similar is it to the corresponding equilibrium phenomenon?

Here we address this question from a theoretical viewpoint.
We study the effect of non-thermal sources of 
energy injection and dissipation on the slow dynamics of glassy 
materials. To obtain quantitative results, we focus 
on a specific, yet 
representative class of glass models whose equilibrium behaviour 
can be studied analytically and is well understood~\cite{KT87},
following exactly the general scenario obtained in the context 
of random first order transition theory~\cite{rfot}. 
Our main result is that 
glass transitions might exist even when driving forces are not thermal. 
Our results also indicate that the location of the transition continuously 
depends on the microsopic details of the dynamics, but that slow 
relaxation near dynamic arrest is insensitive to these details,
thus suggesting that an `effective' equilibrium 
glassy dynamics generically emerges at long times in 
nonequilibrium materials close to dynamic arrest.

To model the experimental situations described above we consider the
dynamics of $N$ degrees of freedom, ${\bf  x} = \{x_i, i=1 \cdots N \}$,
representing for instance the position of grains or cells,
interacting through the Hamiltonian ${\cal H} [{\bf x}]$, 
which is supposed to display a glass transition at thermal 
equilibrium. The driven and active materials we wish to study 
share two important characteristics. First, they dissipate energy
through internal degrees of freedom at a finite rate.
Second, energy is continously supplied either by a global external 
forcing or by the particles themselves. 
To account for these effects, we study the following equation of motion:
\be
\dot{x_i}(t) + \int_{-\infty}^t ds \gamma_d(t-s) 
\dot{x_i} (s) + \frac{\partial {\cal H}}{\partial x_i}
+ \eta_i(t) + f_i^{a}(t) = 0,
\label{eqofmotion}
\ee
where we have included contributions from both a (white noise) equilibrium 
thermal bath 
satisfying the fluctuation-dissipation theorem, 
$\langle \eta_i(t) \eta_j(s) \rangle = 2 T  \delta(t-s) \delta_{ij}$
and from nonequilibrium, colored driving and dissipative mechanisms
represented by the active force $f_i^a(t)$
and the dissipation kernel $\gamma_d(t)$, respectively.
Equation (\ref{eqofmotion})
is a standard theoretical model for the dynamics of active colloids
and molecular motors far from equilibrium~\cite{visco,visco2,blood}. 
It also represents a minimal model to analyze the physics studied
in numerical treatments of active~\cite{mossa,mossa2} and self-propelled 
particles~\cite{marchetti}, where particles 
perform persistent random walks. It certainly misses 
some features of more complicated situations, such as 
complex alignement rules or particle anisotropy~\cite{marchetti2}.

A more intriguing relation, valid when the friction 
(but not the noise) is Markovian, is with 
a Langevin process~\cite{quantum2} that mimics a quantum 
equilibrium bath, and reproduces some features of
the true (operator-valued) quantum Langevin equation~\cite{quantum1}.   
The parameters $\epsilon_a$ and  $\tau_a$ used here are the counterparts 
of $\hbar$ and $\tau_{\rm quant}\equiv \frac{\hbar}{kT}$. 
We shall use this analogy below.

As a first step we choose simple functional forms 
for the colored noise and dissipation 
terms. We use a Gaussian random forcing with mean zero and 
variance $\langle f_i^a (t) f_j^a(s) \rangle = 2 F_a(t-s) \delta_{ij}$,
where
$F_a(t) = \frac{\epsilon_a}{\tau_a}
\exp(- \frac{t}{\tau_a})$,  
with $\tau_a$ the timescale of the slow forcing. 
For the dissipation we similarly choose
$\gamma_d(t) = \frac{\epsilon_d}{\tau_d} \exp(- \frac{t}{\tau_d})$, 
which defines the timescale $\tau_d$.
With these definitions, thermal equilibrium is recovered either 
when $\epsilon_d = \epsilon_a = 0$, or when the colored 
forces and friction satisfy the equilibrium FDT~\cite{kurchan}, 
$F_a(t) = T \gamma_d(t)$. We expect similar results 
for different functional forms for these correlators, as
long as they describe noises with finite correlation times. 

To make the problem tractable, we
perform a mean-field approximation of the glass 
Hamiltonian. Our goal is to have a well-understood 
equilibrium starting point to isolate the influence 
of the non-thermal forces.
We specialize our study to the spherical $p$-spin model, as a well-known 
representative microscopic model where the random first order 
transition scenario becomes exact~\cite{KT87}. It is defined by 
\be
{\cal H} = - \sum_{i_1, \cdots , i_p} J_{i_1 \cdots i_p} x_{i_1} 
\cdots x_{i_p}, 
\label{ham}
\ee
for continuous spins obeying the spherical constraint 
$\sum_i x_i^2 = N$.  
In short, our strategy is to perform a mean-field approximation to 
the interactions in the equation of motion (\ref{eqofmotion}), 
while retaining realistic forms for the sources of injection and 
dissipation. As usual when dealing with glassy dynamics, our
theoretical predictions strictly hold within the particular
context of random first order transition theory, but we expect 
them to have wider physical relevance, see Ref.~\cite{rmp}
for a broad theoretical overview. Another advantage of our 
approach is that it provides precise predictions which
are then useful guides to computer simulations of more realistic 
models of active particles.
 
Because the Hamiltonian (\ref{ham}) is fully-connected, 
closed and exact equations of motion can be derived for 
the autocorrelation function
$C(t,s) = \langle x_i(t) x_i(s) \rangle$, and for the 
autoresponse function $R(t,s) = \partial \langle x_i(t) \rangle
/ \partial \eta_i(s)$:
\begin{widetext}
\begin{eqnarray}
\frac{\partial C(t,s)}{\partial s} & = &
- \mu(t) C(t,s) + \int_{-\infty}^s dt' D(t,t') R(s,t') 
+ \int_{-\infty}^t dt' \Sigma(t,t') C(t',s) 
+ 2 T R(s,t), \nonumber \\
\frac{\partial R(t,s)}{\partial s} & = &
- \mu(t) R(t,s) + \int_{-\infty}^t dt' \Sigma(t,t') R(t',s) 
+ \delta(t-s), \nonumber \\
\mu(t)  & = & T + \int_{-\infty}^t dt' \left[ D(t,t') R(t,t')
+ \Sigma(t,t') C(t,t') \right] ,
\label{detailed}
\end{eqnarray}
\end{widetext}
with the kernels
$D(t,s)  =  \frac{p}{2} C^{p-1}(t,s) + F_a(t-s)$ and 
$\Sigma(t,s) = \frac{p(p-1)}{2}
C^{p-2}(t,s) R(t,s)  + \frac{\partial \gamma_d(t-s)}{\partial s}$. 
The last equation in Eq.~(\ref{detailed}) enforces the spherical constraint.

Technically, introducing colored friction
and noise breaks detailed balance and introduces new physical timescales
($\tau_d$ and $\tau_a$) which compete and perturb the 
equilibrium dynamics of the system. 
This situation is reminiscent of 
the driven dynamics of the model studied in previous work~\cite{BBK}, 
where non-Hamiltonian driving forces were introduced to model 
an applied shear flow. The crucial difference is
the form of the driving terms, whose 
typical timescales in Ref.~\cite{BBK} were that of the dynamics itself, 
while here they relax with their own, 
fixed timescales $\tau_d$ and $\tau_a$ (cf. Eq.~(\ref{detailed})). 
As a result, while the equilibrium glass transition was found to
disappear in the presence of power dissipation of infinitesimally small 
amplitude~\cite{BBK},
we show here that the glass transition may also survive the introduction of 
fluctuating forces, even of large amplitude, a situation which is  
closer to the effect of an oscillating field~\cite{BCI}.

\begin{figure*}
\psfig{file=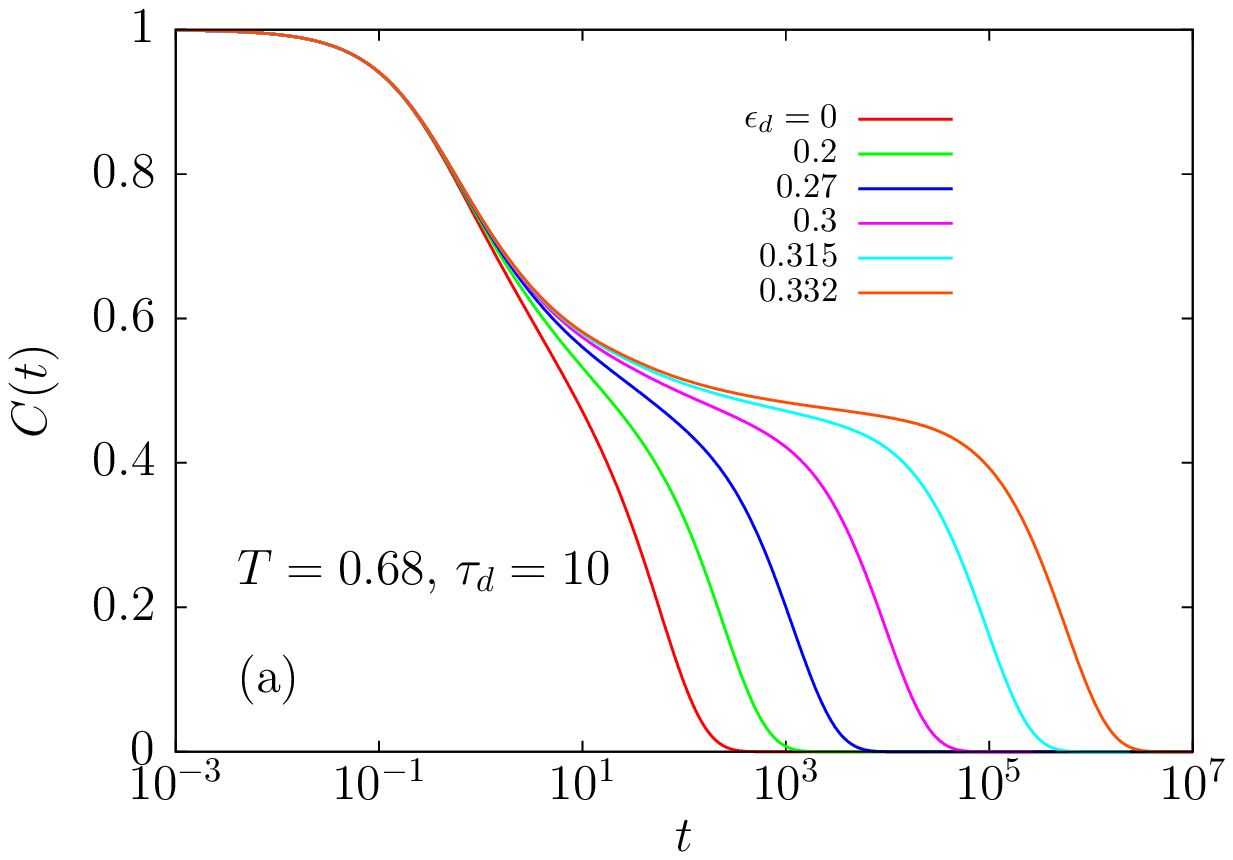,width=5.91cm}
\psfig{file=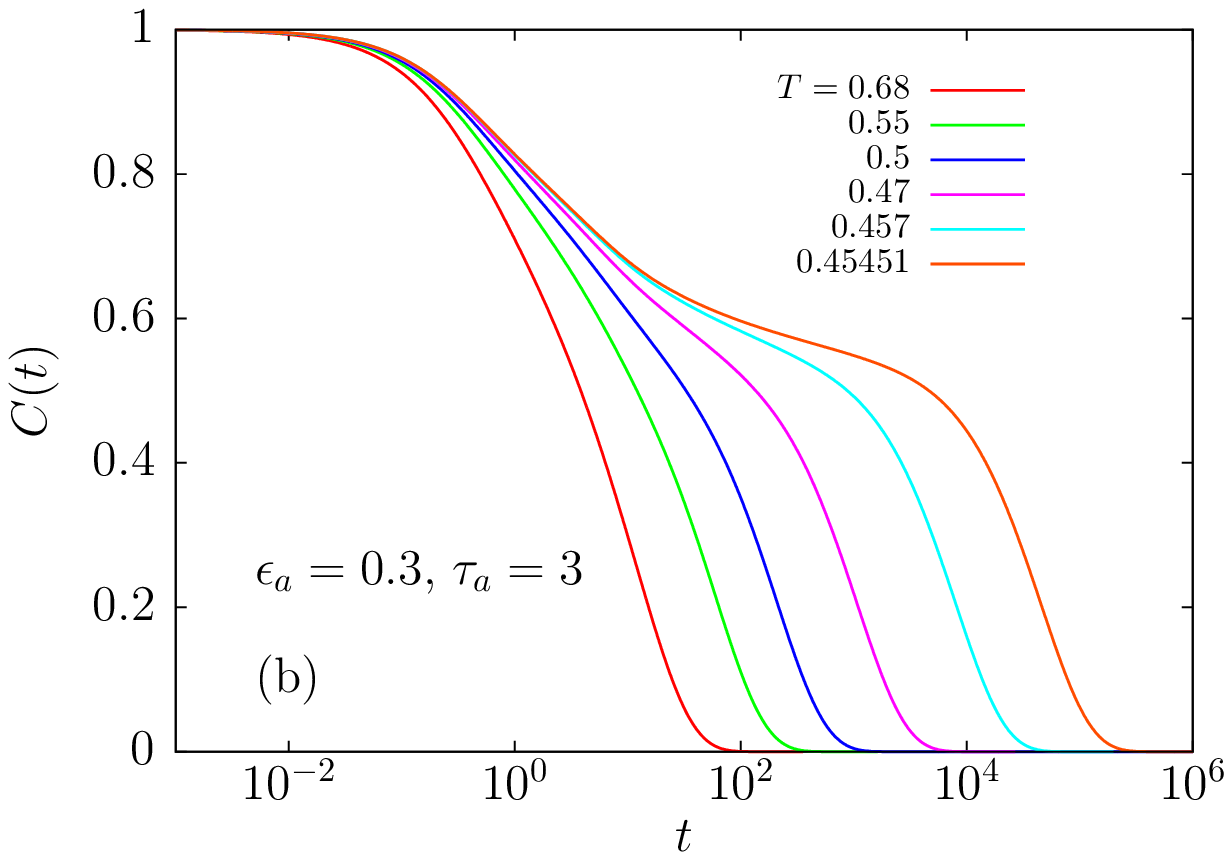,width=5.91cm}
\psfig{file=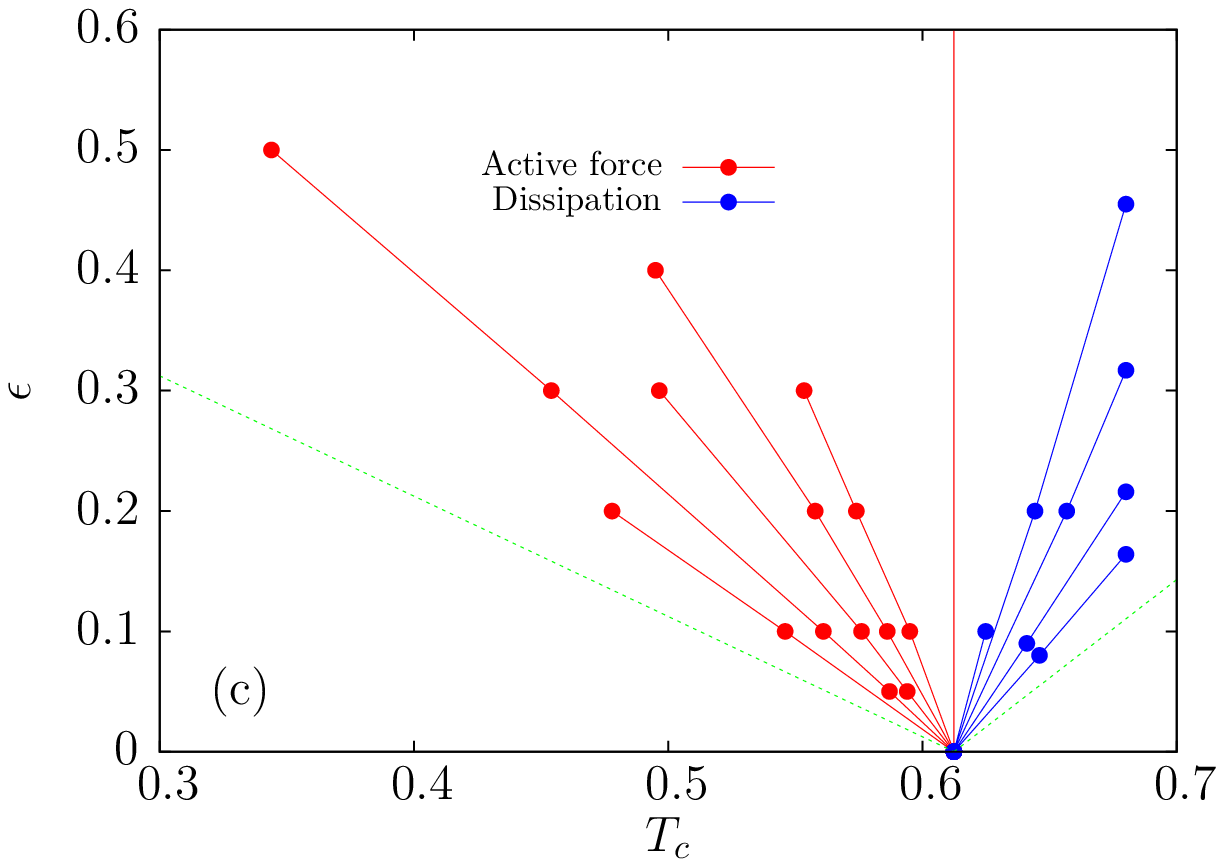,width=5.91cm}
\caption{Slow dynamics with colored dissipation and noise.
(a) Time correlators obtained from the numerical 
solution of Eqs.~(\ref{detailed}) for $p=3$ at 
constant temperature: Increasing the amplitude of the 
dissipation $\epsilon_d$ with $\tau_d=10$ drives the system to 
the glass phase.
(b) Decreasing temperatures for constant 
driving force with $\epsilon_a=0.3$, $\tau_a = 3$.
(c) Phase diagram of the system for active forces 
($T_c < T_c^{\rm eq}$) and slow dissipation ($T_c > T_c^{\rm eq}$), 
for various timescales $\tau_d, \tau_a = 1, 3, 10, 30, 100$ (from bottom
to top). The vertical line indicates $T_c^{\rm eq}$, dashed lines
indicate the limits $\tau_d, \tau_a \to 0$, Eq.~(\ref{barT}).}
\label{corr}
\end{figure*}

The dynamic equations (\ref{detailed}) are 
coupled integro-differential equations for which no general 
analytical solution can be obtained. Our strategy was
to explore the region  in parameter space where 
the dynamics becomes stationary, such that a driven steady state
is reached and the system is in a `fluid' phase with a finite 
relaxation time. We obtained numerical solutions of the 
dynamical equations, which motivated an analytic investigation of their 
asymptotic properties.

We have adapted the numerical scheme
of Ref.~\cite{BBK} to obtain a solution 
of Eqs.~(\ref{detailed}) for a broad range of 
space parameters, using the following values for the 
time scales $\tau_d$ and $\tau_a$: $(1, 3, 10, 30, 100)$, 
in units of the bare friction coefficient which was set 
to unity in Eq.~(\ref{eqofmotion}).
We varied $\epsilon_d$ and $\epsilon_a$ to fully explore the
fluid part of the phase diagram. Numerically, we are limited by the 
stability of the numerical algorithm used to solve  
equations that do not have a formally causal form in 
the time differences~\cite{BBK}.
We illustrate our results in Fig.~\ref{corr} with the 
solution obtained in two limiting cases where 
either only dissipation or active driving forces  are 
introduced. We do not find any added complexity when both terms 
are simultaneously present with different amplitudes and timescales. 

A direct analytical solution only 
exists in the trivial limit where
$\tau_d, \tau_a \to 0$, because Eq.~(\ref{eqofmotion})
reduces to a standard Langevin dynamics with 
white noise and memoryless friction. Equilibrium
is then achieved at the rescaled temperature
\be 
\bar{T} = \frac{T+\epsilon_a}{1+\epsilon_d}.
\label{barT}
\ee 
Below, we argue that for finite 
correlation times of driving and dissipation forces, 
a (more complicated) relation to the equilibrium situation {\it only 
holds in the limit of very large time scales}, 
a result which is far more subtle than the mapping 
in Eq.~(\ref{barT}).
Equation (\ref{barT}) implies
in particular that response and correlation functions satisfy 
the fluctuation-dissipation theorem, ${\bar T} R(t) = - \frac{dC(t)}{dt}$, 
and that 
Eqs.~(\ref{detailed}) reduce to 
\be 
\frac{dC(t)}{dt} + {\bar T} C(t) 
- \frac{p}{2 {\bar T}} \int_0^t dt' C^{p-1}(t-t') \frac{dC(t')}{dt'} = 0,
\label{equil}
\ee  
which is mathematically equivalent~\cite{KT87} to the
so-called $F_{p-1}$ schematic model derived in the context 
of the mode-coupling theory (MCT) of the glass transition~\cite{gotze}. 
Its solution is known in great detail, and
displays a dynamic singularity as the (rescaled) temperature 
is lowered towards the equilibrium value $T_c^{\rm eq}$. Near the dynamic 
transition, the correlation function develops a two-step decay, 
with asymptotic time dependences that follow the 
behaviour described for discontinuous (or `type B') transitions 
within MCT~\cite{gotze}. An expression for the critical temperature 
and scaling laws is obtained by performing a detailed mathematical 
analysis~\cite{gotze} of the equation derived by 
taking the long-time limit of 
Eq.~(\ref{equil}),
\be
{\bar T} C_s(t) 
+ \frac{1}{{\bar T}} \int_0^t dt' D_s(t-t') \frac{dC_s(t')}{dt'} 
- \frac{1-q}{\bar T} D_s(t) = 0,
\label{equils}
\ee 
where $D_s(t) = \frac{p}{2} C_s^{p-1}(t)$.

Note that, in the general driven case with a nonequilibrium bath,
one can only go from Eqs.~(\ref{detailed}) to Eq.~(\ref{equil}) by formally 
eliminating  $R(t-t')$ from the dynamic equations.
In this way, one only obtains a version of Eq.~(\ref{equil}) with the 
memory kernel taking a  form ${\cal M}[C](t-t')$, which is a complicated 
{\it nonlocal} functional of $C(t)$ (as opposed to the local
functions like $C^{p-1}(t)$).

Our central result is that the main features of this equilibrium 
glass transition robustly survive the introduction of a finite 
amount of non-thermal fluctuations driving the system far from thermal
equilibrium, as illustrated in Fig.~\ref{corr}. In both cases 
of colored noise or dissipation, we numerically find that time 
correlation functions display a two-step decay reminiscent 
of the equilibrium behaviour. This emerging glassy
dynamics is characterized by a relaxation time 
that diverges upon approaching a
dynamic transition, which we call a {\it nonequilibrium glass transition}. 
We find that the location of the transition 
is a continuous function of the driving mechanism. 
The data in Fig.~\ref{corr} confirm the natural expectation 
that $T_c$ increases in the presence of the 
additional colored dissipation, 
$T_c(\epsilon_d, \tau_d) > T_c^{\rm eq}$, 
while it decreases in the presence
of a colored forcing, 
$T_c(\epsilon_a, \tau_a) < T_c^{\rm eq}$. The phase diagram 
determined numerically in Fig.~\ref{corr}c
suggests a nearly linear dependence of $T_c$ on either 
$\epsilon_a$ or $\epsilon_d$, with a less pronounced dependence on the 
time scales (note that $\tau_d$ and $\tau_d$ vary 
over 2 orders of magnitude in Fig.~\ref{corr}c).  

\begin{figure*}
\psfig{file=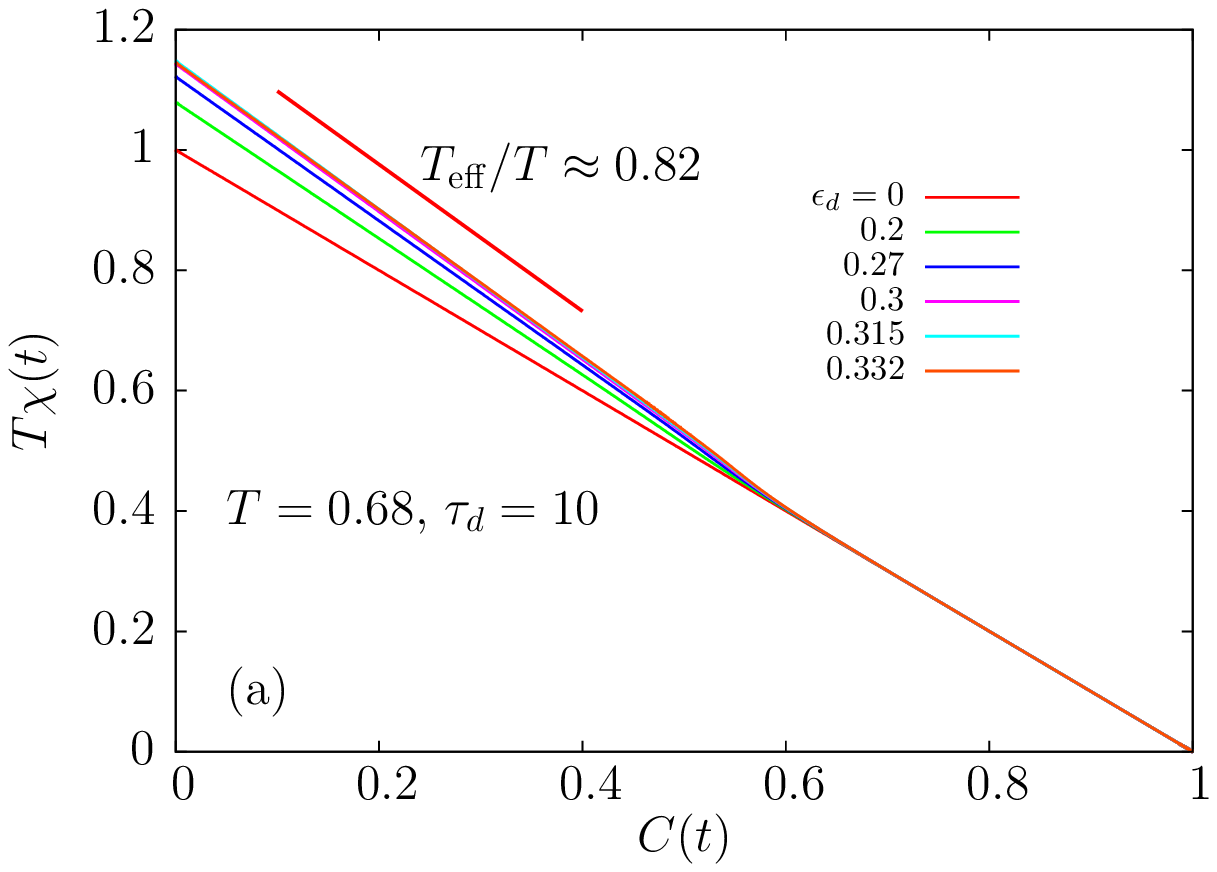,width=5.91cm}
\psfig{file=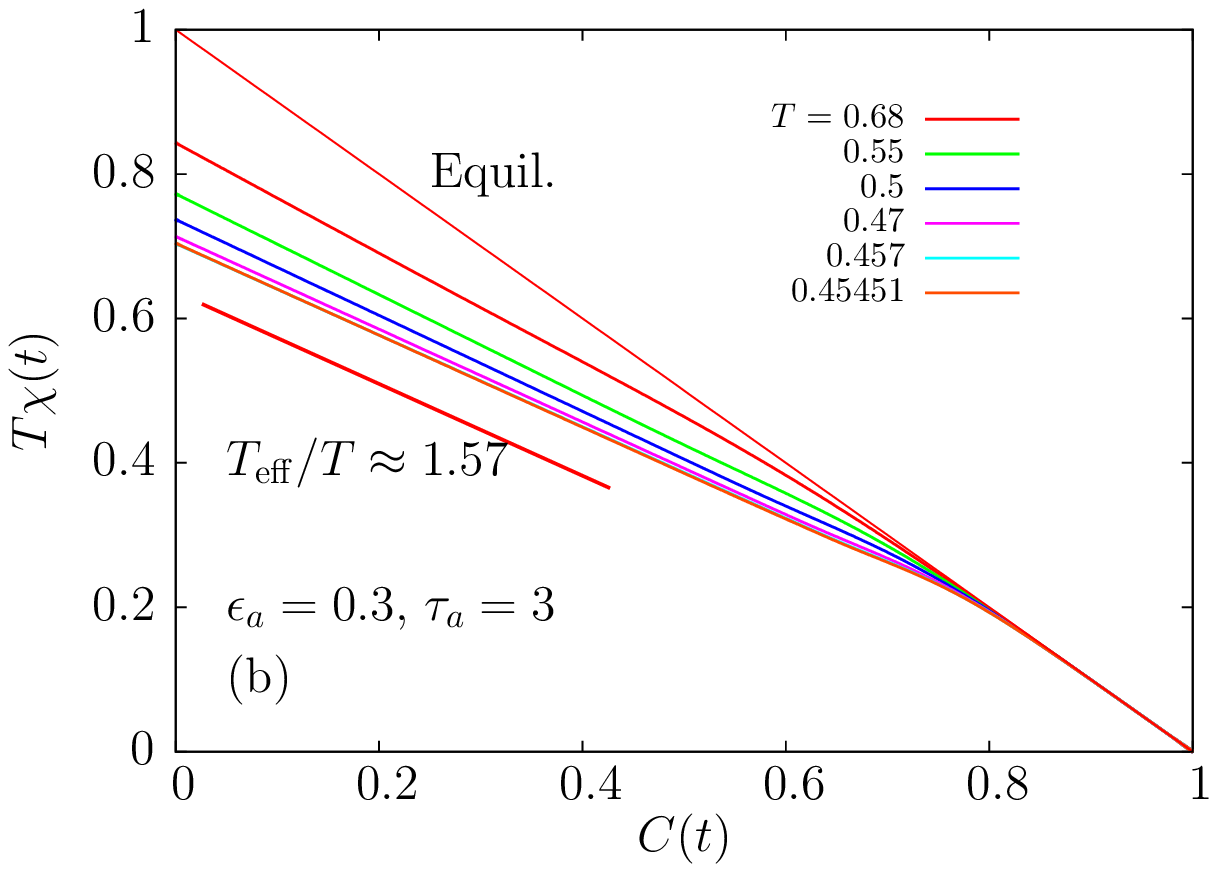,width=5.91cm}
\psfig{file=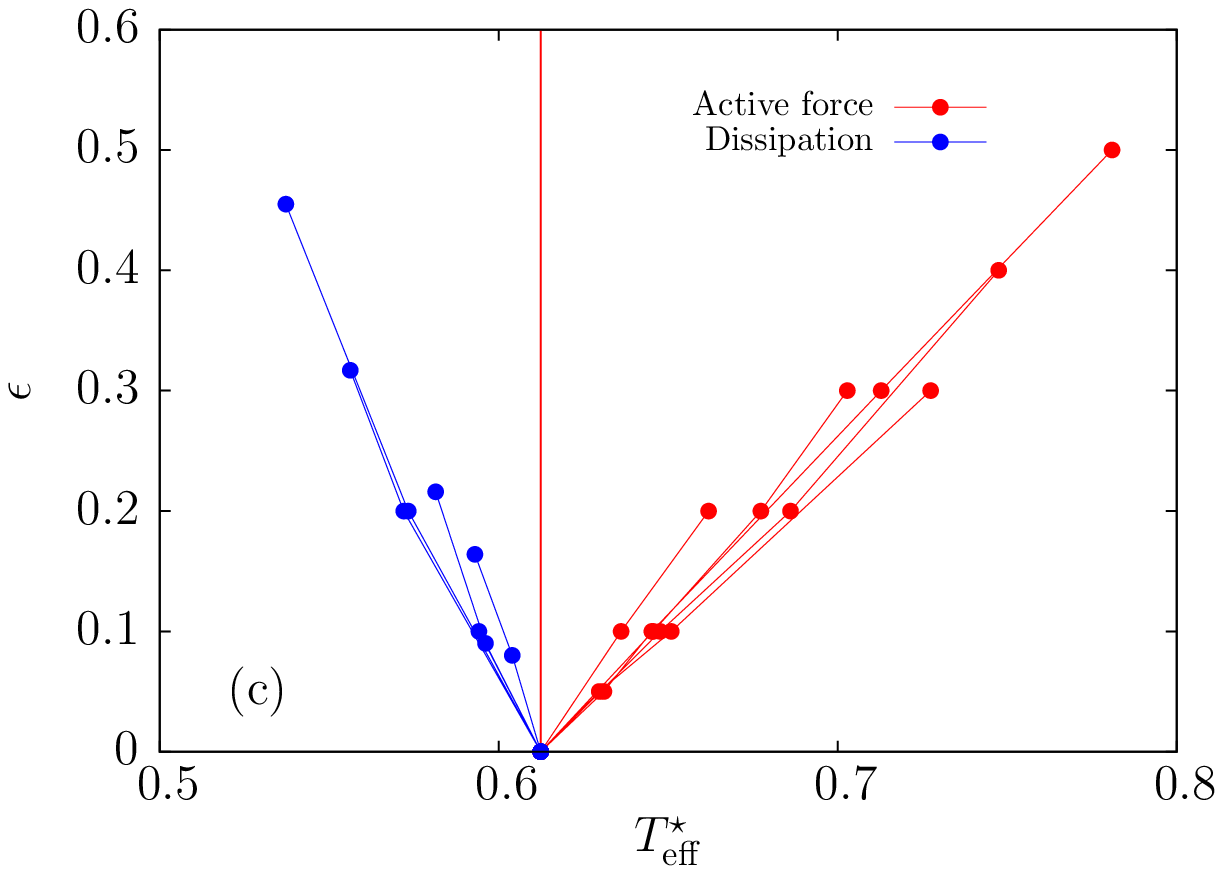,width=5.91cm}
\caption{Effective temperatures near nonequilibrium 
glass transitions.
(a) Dissipative forces produce $T_{\rm eff} < T$ (i.e. $X_s > 1$).  
Parameters as in Fig.~\ref{corr}a. 
(b) Active driving produces $T_{\rm eff} > T$ (i.e. $X_s < 1$).
Parameters as in Fig.~\ref{corr}b.
(c) Effective critical temperatures for parameters as in
Fig.~\ref{corr}c.}   
\label{phase}
\end{figure*}

We can analytically rationalize the above findings, and obtain 
additional insight into the slow dynamics near nonequilibrium 
glass transitions, relying on the fact that the driving terms 
responsible for the explicit violation of detailed balance
in the dynamical equations (\ref{detailed}) have correlations 
that vanish at long times. Thus, a 
strong scale separation occurs
when the structural relaxation is much
larger than both $\tau_d$ and $\tau_a$, i.e. 
sufficiently close to the nonequilibrium transition. 
We seek an approximate equation of motion valid 
for stationary states in the limit of
large times, $\tau_d, \tau_a \ll t$, corresponding to 
the approach and departure from the plateau: 
\begin{widetext}
\begin{eqnarray}
\frac{\partial C_s(t)}{\partial t} & = & - ( \mu - I_\Sigma) C_s(t)
+ I_R D_s(t) 
+ \int_0^t dt' \Sigma_s(t-t') C_s(t')
+ \int_0^\infty dt' \left[ 
D_s(t+t') R_s(t') + \Sigma_s(t+t') C_s(t') \right],
\label{slowcorr}  \\
\frac{\partial R_s(t)}{\partial t} & = & - (\mu -I_\Sigma) R_s(t)
+ \Sigma_s(t) I_R
+ \int_0^t dt' \Sigma_s(t-t') R_s(t'), \label{slowresp} \\
\mu & = & T + \Omega + \int_0^\infty dt' \left[ 
D_s(t') R_s(t') + \Sigma_s(t') C_s(t') \right],
\label{slowmu}
\end{eqnarray}
\end{widetext}
with 
$\Sigma_s = \frac{p(p-1)}{2} C_s^{p-2} R_s$, 
$D_s = \frac{p}{2} C_s^{p-1}$, 
and the following integrals
$I_\Sigma = \int_0^\infty dt' \Sigma_f(t')$, 
$I_R = \int_0^\infty dt' R_f(t')$, 
and $\Omega = \int_0^\infty dt' \left[ 
D_f(t') R_f(t') + \Sigma_f(t') C_f(t') \right]$.
We defined the `slow' functions $C_s(t)$ and $R_s(t)$ 
as the exact solutions of Eqs.~(\ref{slowcorr}-\ref{slowmu}), 
while the `fast' ones are defined by difference, e.g.
$C_f (t) = C(t) - C_s(t)$, and decay over time scales
that do not diverge at the transition. 
 
A crucial element of the dynamical equations (\ref{slowcorr}-\ref{slowmu}) 
governing the long-time dynamics is that the terms 
responsible for the explicit breaking of detailed balance
have disappeared. They appear very indirectly through 
time integrals over the short-time dynamical behaviour.
As an immediate  consequence, these equations can be considerably 
simplified because Eqs.~(\ref{slowcorr}, \ref{slowresp}) reduce to 
the same equation if correlation and response satisfy  
\be
R_s(t) = - \frac{X_s}{T} \frac{dC_s(t)}{dt},
\label{ansatz}
\ee
which defines the fluctuation-dissipation ratio, $X_s$, 
or equivalently an effective temperature $T_{\rm eff} = T/X_s$~\cite{CKP97}.
A similar ansatz holds in the long-time limit of the aging 
regime~\cite{cuku93}, i.e. in the unperturbed glass phase, and in the 
equivalent limit of vanishing shear-like forces~\cite{BBK}. 

Combining Eqs.~(\ref{slowcorr}, \ref{slowmu}, 
\ref{ansatz}), we obtain
\begin{eqnarray} 
& & ( T + \frac{p}{2} X_s q^p + \Omega - I_\Sigma 
- \frac{p}{2} q^{p-1} ) C_s(t) 
\nonumber \\
& & 
  + \frac{X_s}{T} \int_0^t dt' D_s(t-t') \frac{d C_s(t')}{dt'} 
- I_R D_s(t) = 0 
\label{final}
\end{eqnarray}
where $q$ represents the intermediate plateau height of $C(t)$. 
This equation constitutes our main analytical achievement. 
It shows that, at sufficiently long times, the dynamical equation governing 
structural relaxation is equivalent to the 
one found for equilibrium relaxation, Eqs.~(\ref{equil}, \ref{equils}),
showing that an equilibrium-like glassy dynamics emerges out 
of nonthermal forces driving the dynamics at short times.
A similar conclusion was reached in Ref.~\cite{zippelius} for the specific 
case a driven granular fluid, but {\it here we find that this form 
only holds once the fast motion is averaged away}.
Even then, there remain, however, two important differences 
with the equilibrium case.

First, the `coupling' parameters determining the numerical value 
of the critical temperature are `renormalized' by the microscopic details
of the driving forces through time integrals over the short-time 
dynamics, as can be seen by directly 
comparing Eqs.~(\ref{equils}) and (\ref{final}).
This explains our numerical finding 
that the location of the transition 
and the value of the plateau height
continuously depend on the details of the microscopic dynamics, 
Fig.~\ref{corr}c. Thus, an analytic 
determination of the locus of the nonequilibrium glass transition 
requires solving not only Eq.~(\ref{final}), but also 
strongly nonuniversal features of the short-time dynamics. 
By contrast, because
the long-time dynamics remains described by a discontinuous MCT 
transition, all universal features of the time correlation 
functions remain valid far from equilibrium, as
found numerically in Fig.~\ref{corr}. In particular, while the 
critical exponents describing the time dependence of correlation 
functions will depend on the driving forces, the well-known 
relations between them~\cite{gotze} remain valid.

Second, while Eq.~(\ref{final}) simply involves the 
correlator $C_s(t)$, as in equilibrium, the 
response function $R_s(t)$ does not obey the  
equilibrium fluctuation-dissipation relation, but only 
an `effective' one, Eq.~(\ref{ansatz}). Note that we 
predict the existence of a nonequilibrium $T_{\rm eff}$ 
for slow degrees of freedom even for the 
stationary fluid phase, not only deep into the glass as in 
Ref.~\cite{trieste}.
This finding illustrates that nonequilibrium glass transitions 
are conceptually distinct from the equilibrium analog, 
but that a form of equilibrium-like glassy dynamics 
naturally emerges at long times.

Our numerical analysis confirms the existence of effective temperatures, 
see Fig.~\ref{phase}.
As usual~\cite{cuku93,CKP97}, we represent 
integrated response functions, $\chi(t) = \int_0^t dt' R(t')$,
versus $C(t)$ using $t$ as a running parameter, which yields 
straight lines with slope $-X_s/T$ whenever Eq.~(\ref{ansatz}) holds. 
The straight segments corresponding  to the first step of the relaxation 
have no reason, despite appearances, to be straight since 
the system is strongly driven.
The data in Fig.~\ref{phase} show that $X_s$ 
behaves differently if friction ($X_s (\epsilon_d , \tau_d) > 1$),
or forcing ($X_s (\epsilon_a , \tau_a) < 1$)
dominates the physics. The former represents an unusual situation
where slow degrees of freedom appear to be colder than 
the bath~\cite{peter}. Although Eq.~(\ref{barT}) cannot 
be used to predict the actual value of $T_{\rm eff}$ in the general
case, it correctly predicts its qualitative trends and thus 
provides a simple physical argument for its variation with our 
control parameters.
These two distinct cases are 
reminiscent of the distinction between adamant (`hot') 
and susceptible (`cold') molecular motors of 
Ref.~\cite{wolynes}.  Note also the striking similarity with 
the corresponding curves measured in active fluids~\cite{mossa,mossa2}.

Finally, we summarize our numerical findings for 
$X_s$ measured at the dynamic transition into a
`phase diagram' ($\epsilon, T_{\rm eff}^\star = T_c/X_s$) in 
Fig.~\ref{phase}c.  With friction alone, we confirm that 
$T_{\rm eff}^\star < T_c(\epsilon_d,\tau_d)$ (i.e. $X_s > 1$), but the resulting 
$T_{\rm eff}^\star$ is lower than the equilibrium value $T_c^{\rm eq}$.
The additional dissipation thus drives the transition 
temperature $T_c$ up, but it shifts the `effective' critical 
temperature $T_{\rm eff}^\star$ down. 
Active forces have the opposite effect, 
see Figs.~\ref{phase}b,c. Somewhat disappointingly, 
the physical intuition~\cite{wolynes} that nonequilibrium transitions occur 
at shifted $T_c$ but constant $T_{\rm eff}^\star 
\approx T_c^{\rm eq}$ is {\it not} valid for our simple model: 
the effective temperature cannot be used to infer the nonequilibrium
phase diagram in our model. This can also be seen analytically,
as the mapping between equilibrium and off-equilibrium 
dynamics is more complicated than simply replacing $T$ by $T_{\rm eff}$,
see Eqs.~(\ref{equils}, \ref{final}).

Let us discuss the general picture of the $(T,\epsilon)$ phase 
diagram, Fig. 1c, in particular for the active forces. 
The red lines mark a limit between a liquid phase (higher $T$ 
and $\epsilon$) and a glass phase where the dynamics 
performs aging. Our numerical method does not allow us to 
determine whether this transition line continues up to $T=0$,
as the numerical solution of the equations becomes unstable.  
One possibility is that the red lines continue and
incide at the $T=0$ axis without qualitative changes, thus enclosing 
the glassy phase. An alternative is suggested 
by the close analogy with the quantum Langevin equation described above.
For the case of the $(T,\hbar)$ plane, it is well established~\cite{CuGr} 
that a continuous transition line (very much like our
red lines) exists up to a point where the transition becomes 
thermodynamically first order all the way to $T=0$ 
(see Fig.~2 of Ref.~\cite{CuGr}).
For our model, it would mean that the transition 
crossed by decreasing $\epsilon$ at constant (very) low temperature
is first order: the system would vitrify suddenly from a liquid 
phase without any precursor increase in viscosity. Such an intriguing 
phenomenon, seemingly consistent with the data in Ref.~\cite{marchetti},
clearly deserves further study.  
 
Our study provides firm theoretical grounds to 
the emerging view that dynamic arrest in driven and active materials 
shares important similarities with the equilibrium glass transition, 
in particular regarding the behaviour of time 
correlation functions and the emergence of an `effective' thermal 
behaviour. Since our study was performed within a particular 
theoretical framework, it would be interesting
to check the robustness of our findings
beyond the realm of schematic glassy models,
for instance using nonequilibrium mode-coupling 
types of approximations~\cite{chorin,brader}. 
Further away from mean-field approximations, one could  
for instance also test in actual measurements 
whether the role of `activated processes'~\cite{rmp} 
remains the same in and out of equilibrium, and if the
mode-coupling crossover is as relevant in driven 
systems as it is in equilibrium ones.

Regarding effective temperatures, 
a numerical study of a kicked granular fluid~\cite{zippelius} reports 
a shift of the glass transition towards larger density with 
increasing activity, suggesting that drive dominates 
over friction in that case. Thus, we predict the existence of 
an effective temperature $T_{\rm eff} > T$ for this system.
Such a result is reported in numerical studies 
of active particle systems~\cite{mossa,mossa2}, but
these measurements were not performed close to dynamic arrest. 
Closer to our ideas are the experimental measurements
of an effective temperature in an air-fluidized granular 
bed~\cite{durian2}, then used to infer a nonequilibrium equation of state. The 
reported deviations from equilibrium behaviour~\cite{durian2} are highly 
reminiscent of our findings in Fig.~\ref{phase}.
We believe that future studies of active particles 
at large density~\cite{marchetti},
or biologically driven systems~\cite{blood} will result in
more examples of nonequilibrium glass transitions 
as described in this work, where tests of 
our predictions about the universality of the slow dynamics 
and the emergence of effective temperatures could be 
performed. 

{\bf Acknowledgements:}
We thank E. Bertin and O. Dauchot for discussions, 
Groupement de Recherches PHENIX and  ANR programme JAMVIB
for partial financial supports. 
The research leading to these results has received funding
from the European Research Council under the European Union's Seventh
Framework Programme (FP7/2007-2013) / ERC Grant agreement No 306845.

{\bf Author contributions:}
Both authors contributed equally to this work.

\end{document}